\documentclass[fleqn,usenatbib]{mnras}

\usepackage{newtxtext}
\usepackage[varvw]{newtxmath}

\usepackage[T1]{fontenc}

\DeclareRobustCommand{\VAN}[3]{#2}
\let\VANthebibliography\thebibliography
\def\thebibliography{\DeclareRobustCommand{\VAN}[3]{##3}\VANthebibliography}

\usepackage{diffcoeff}
\usepackage{cleveref}
\usepackage{float}
\usepackage{placeins}

\usepackage{scalerel}
\usepackage{tikz}
\usetikzlibrary{svg.path}

\definecolor{orcidlogocol}{HTML}{A6CE39}
\tikzset{
  orcidlogo/.pic={
    \fill[orcidlogocol] svg{M256,128c0,70.7-57.3,128-128,128C57.3,256,0,198.7,0,128C0,57.3,57.3,0,128,0C198.7,0,256,57.3,256,128z};
    \fill[white] svg{M86.3,186.2H70.9V79.1h15.4v48.4V186.2z}
                 svg{M108.9,79.1h41.6c39.6,0,57,28.3,57,53.6c0,27.5-21.5,53.6-56.8,53.6h-41.8V79.1z M124.3,172.4h24.5c34.9,0,42.9-26.5,42.9-39.7c0-21.5-13.7-39.7-43.7-39.7h-23.7V172.4z}
                 svg{M88.7,56.8c0,5.5-4.5,10.1-10.1,10.1c-5.6,0-10.1-4.6-10.1-10.1c0-5.6,4.5-10.1,10.1-10.1C84.2,46.7,88.7,51.3,88.7,56.8z};
  }
}

\newcommand\orcidicon[1]{\href{https://orcid.org/#1}{\mbox{\scalerel*{
\begin{tikzpicture}[yscale=-1,transform shape]
\pic{orcidlogo};
\end{tikzpicture}
}{|}}}}

\usepackage{xcolor}
\newcommand\hl[1]{%
  \bgroup
  \hskip0pt\color{red!80!black}%
  #1%
  \egroup
}

\usepackage{hyperref}

\usepackage{graphicx}	
\usepackage{amsmath}	

\title[Enhanced Phase Mixing of Torsional Alfv\'en Waves]{Enhanced Phase Mixing of Torsional Alfv\'en Waves in Stratified and Divergent Solar Coronal Structures, Paper II: Nonlinear Simulations}

\author[C. Boocock and D. Tsiklauri]{
C. Boocock$^1$\thanks{E-mail: c.boocock@qmul.ac.uk} \orcidicon{0000-0002-8080-0555} and D. Tsiklauri$^2$ \orcidicon{0000-0001-9180-4773}
\\
$^1$Queen Mary University of London, Mile End Road, London E1 4NS, UK, \\
$^2$School of Science and Technology,
University of Georgia,
77a Kostava Street,
0171 Tbilisi,
Georgia
}

\date{Accepted 2021 December 6. Received 2021 December 2; in original form 2021 October 27}

\pubyear{2021}

\begin{document}
\label{firstpage}
\pagerange{\pageref{firstpage}--\pageref{lastpage}}
\maketitle

\begin{abstract}
We use MHD simulations to detect the nonlinear effects of torsional Alfv\'en wave propagation in a potential magnetic field with exponentially divergent field lines, embedded in a stratified solar corona. In Paper I we considered solutions to the linearised governing equations for torsional Alfv\'en wave propagation and showed, using a finite difference solver we developed named \textit{WiggleWave}, that in certain scenarios wave damping is stronger than what would be predicted by our analytic solutions. In this paper we consider whether damping would be further enhanced by the presence of nonlinear effects.  We begin by deriving the nonlinear governing equations for torsional Alfv\'en wave propagation and identifying the terms that cause coupling to magnetosonic perturbations. We then compare simulation outputs from an MHD solver called \textit{Lare3d}, which solves the full set of nonlinear MHD equations, to the outputs from \textit{WiggleWave} to detect nonlinear effects such as: the excitation of magnetosonic waves by the Alfv\'en wave, self-interaction of the Alfv\'en wave through coupling to the induced magnetosonic waves, and the formation of shock waves higher in the atmosphere caused by the steepening of these compressive perturbations. We suggest that the presence of these nonlinear effects in the solar corona would lead to Alfv\'en wave heating that exceeds the expectation from the phase mixing alone.
\end{abstract}




\section{Introduction}
\label{sec:intro}

In this paper we consider the propagation and enhanced phase mixing of torsional Alfv\'en waves in open structures in the solar corona with exponentially diverging magnetic field lines and a gravitationally stratified atmosphere. For a long time wave-based heating has been considered as a mechanism to heat the solar corona \citep{Ofman_holes}. A large variety of magnetic waves have been detected in the corona \citep{Parnell}. In particular torsional Alfv\'en waves have recently been detected in the photosphere \citep{Stangalini2021} and it has been shown in \citep{Soler} that torsional Alfv\'en waves of intermediate frequencies are able to penetrate to the corona, and that expanding flux tubes favour this transmission. Torsional Alfv\'en waves can also be generated in the corona when the magnetic field relaxes following a solar flare \citep{FletcherHudson}.

Although Alfv\'en waves are viable transporters of non-thermal energy to the solar corona they are difficult to dissipate due to the high conductivity of the corona, \citep{Jigsaw}. Dissipation must therefore occur over small length scales typically generated by in inhomogeneities in the corona, \citep{DM2008}. Phase mixing is a mechanism first proposed by \citep{Priest} that increases the viscous and ohmic dissipation of Alfv\'en waves in the solar corona. When Alfv\'en waves on neighboring magnetic surfaces propagate at different speeds they move out of phase with one another leading to strong gradients transverse to the direction of propagation, thus increasing the effect of viscous or resistive heating in the plasma.

Enhanced phase mixing occurs in divergent magnetic structures such as those that are found in open field regions or at the base of large (50-100 Mm height) coronal loops. The decrease of magnetic field strength along field lines means a decreasing Alfv\'en velocity that in turn reduces the wavelength of propagating Alfv\'en waves and ultimately enhances the effect of phase mixing. Conversely a stratified density gradient has the opposite effect, by increasing the Alfv\'en velocity in the direction of propagation, wavelengths are increased and the effect of phase mixing is diminished \citep{Smith}.

In Paper I we considered solutions to the linearised governing equations for the propagation of torsional Alfv\'en waves in a potential magnetic field with exponentially divergent field lines, embedded in a stratified solar corona. We used the WKB approximation to derive an analytic solution for Alfv\'en wave propagation and damping due to enhanced phase mixing. We also developed an IDL code, \textit{TAWAS} (\url{https://github.com/calboo/TAWAS}), to calculate this solution over a coordinate grid. We then developed a finite difference solver \textit{Wigglewave} (\url{https://github.com/calboo/Wigglewave}) that solves the linearised governing equations directly. 

By comparing the outputs from these two codes we showed that the analytic solution is accurate within the limits of the WKB approximation, but beyond the limits of the WKB approximation the analytic formula under-reports the effect of Alfv\'en wave damping. We also showed that torsional Alfv\'en waves driven from the base of the corona can be fully dissipated within 100 Mm if the field is magnetic field lines are highly divergent, the wave period is low and a high value of anomalous kinematic viscosity is applied. These conditions could be relaxed if additional physical effects are considered such as pressure, three-dimensionality and nonlinearity \citep{Smith}. In this paper we consider the effects of nonlinearity on the propagation and enhanced phase mixing of torsional Alfv\'en waves. We do this by using an MHD solver called \textit{Lare3d}, \citep{lare3d}, to simulate torsional Alfv\'en wave propagation using the full nonlinear MHD equations.

In order to detect the effects nonlinearity we compare the outputs from \textit{Lare3d} to those from \textit{Wigglewave}. We also compare the outputs from several \textit{Lare3d} simulations with different amplitudes for the driving Alfv\'en wave. We are able to detect effects such as the excitation of magnetosonic waves by Alfv\'en wave phase mixing, as described in \citep{Nakariakov1997} and simulated in \citep{Shestov} who demonstrate the excitation of both fast and slow modes; the trapping of magnetosonic waves due to refraction within the tube structure, as mentioned in \citep{Roberts}; the generation of compressive perturbations associated with the magnetosonic waves, as identified in \citep{Ofman} and the steepening of these perturbations into shock waves as discussed in \citep{Malara_Compressible}.

Although we are unable to directly measure Alfv\'en wave damping from our \textit{Lare3d} simulation outputs we can infer from previous research that the emergence of nonlinear effects, such as coupling to magnetosonic waves and shock formation, provide an alternative, and often more efficient, mechanism for Alfv\'en wave dissipation and viscous heating \citep{Arber}, \citep{Malara_3d}, \citep{Malara_Compressible}.

In \cref{sec:nonlin} we derive the nonlinear equations for the propagation of torsional Alfv\'en waves in an axisymmetric potential magnetic field. Using these we can see how the linearised equations were originally derived and we can identify the nonlinear terms that give rise to additional physical effects. In \cref{sec:simulations} we will see these effects arising in 3D MHD simulations performed using \textit{Lare3d}.

\section{Nonlinear effects of Torsional Alfv\'en waves propagating in a Potential Axisymmetric Magnetic Field}
\label{sec:nonlin}
 
 In this section we will show mathematically which terms in the nonlinear MHD equations cause the system to deviate from purely torsional and incompressible perturbations and which physical effects these terms correspond to. To do this we derive the full nonlinear equations for torsional Alfv\'en waves propagating in a potential axisymmetric magnetic field and identify which terms are nonlinear and how they effect the system.
 
 We demonstrate how torsional Alfv\'en waves propagating in a potential axisymmetric magnetic field can excite magnetosonic waves which can then interact back with the Alfv\'en waves and cause perturbations to the density. We begin with the cold $(\beta = 0)$, ideal (zero viscosity, zero resistivity) MHD equations,

\begin{equation}
\begin{split}
\rho\frac{\partial\mathbf{v}}{\partial t}  + \rho(\mathbf{v}\cdot\nabla)\mathbf{v} \quad= \quad & -\frac{1}{\mu_{0}}\mathbf{B}\times(\nabla\times\mathbf{B}),
\end{split}
\end{equation}
\begin{equation} 
\begin{split}
\frac{\partial\mathbf{B}}{\partial t}  \quad= \quad & \nabla \times (\mathbf{v} \times \mathbf{B}), \end{split}
\end{equation}
\begin{equation} 
\begin{split}
\frac{\partial\rho}{\partial t} + \nabla\cdot(\rho\mathbf{v}) \quad = \quad & 0, \\
\end{split}
\end{equation}
\begin{equation}
\nabla\cdot\mathbf{B}  \quad= \quad 0.  \\
\end{equation}

We have omitted the inclusion of viscosity in the momentum equation for conciseness and simplicity, furthermore as the kinematic viscosity $\nu$ is constant, the viscosity term, $\nu\nabla^2v$, is linear in $v$ so would be unaffected by the eventual linearisation of these equations. We consider these MHD equations in cylindrical coordinates $(r,z,\theta)$ and with the condition of axisymmetry ($\partial/\partial \theta =0$), the resultant components of the MHD equations are shown in \cref{app1}. We then consider perturbations to an equilibrium state, $\mathbf{v_0} = \mathbf{0}$, $\mathbf{B_0} = (B_{0r},0,B_{0z})$ and $\rho_0 = \rho_0(r,z)$, that take the form, 

\begin{equation}
\begin{split}
\rho \quad = & \quad \rho_0 + \rho', \\
\mathbf{v} \quad = & \quad \mathbf{0} + \mathbf{v},\\
\mathbf{B} \quad = & \quad \mathbf{B_0} + \mathbf{b}, \\
\end{split}
\end{equation}
where our equilibrium magnetic field is a potential field such that,

\begin{equation}
\nabla\times\mathbf{B_0} = \begin{pmatrix}0\\\partial_zB_{0r}-\partial_rB_{0z}\\0\end{pmatrix} =\mathbf{0}.
\end{equation}

The resultant nonlinear equations for these perturbations are also given in \cref{app1} with the linear and nonlinear terms shown separately. Looking at these equations, we can associate the variables $v_\theta$ and $b_\theta$ with Alfv\'en waves and the variables $v_r,v_z,b_r,b_z$ with magnetosonic waves.

We can see that the nonlinear terms, which correspond to the evolution equations for $v_r,v_z,b_r,b_z$, contain terms with either two magnetosonic variables or two Alfv\'en wave variables, indicating that the magnetosonic waves can be excited by Alfv\'en waves and can self-interact. 

In contrast the nonlinear terms that correspond to the evolution equations for $v_\theta$ and $b_\theta$, contain only terms with an Alfv\'en wave variable and a magnetosonic variable, indicating that they cannot be excited by magnetosonic waves but can interact with the induced magnetosonic waves. In other words, the torsional Alfv\'en waves can only self-interact in the presence of magnetosonic perturbations \citep{Farahani}.

If the nonlinear terms in these equations are ignored, and we imagine a point in time where perturbations are purely torsional, such that $v_r,b_r,v_z,b_z$ and $\rho' = 0$, then the system stays incompressible and the perturbations remain purely torsional. The equations for this system are given in \cref{app2}. By simplifying the equations for $v_\theta$ and $b_\theta$ and reintroducing the viscosity term, we arrive back at the linearised governing equations for torsional Alfv\'en wave propagation presented in Paper I and shown below in \cref{eq:velocity,eq:magnetic}, this transformation is shown explicitly in \cref{app2}.

\begin{equation}
\label{eq:velocity}
    \rho \diffp {v}{t}  = \quad 
    \frac{1}{r\mu_{0}}\left(\mathbf{B_0}\cdot\nabla(rb)\right)
    + \frac{1}{r}\diffp{}{r}\left(\rho\nu r\diffp{v}{r}\right)
    + \diffp{}{z}\left(\rho\nu\diffp{v}{z}\right),
\end{equation}
\begin{equation}
\label{eq:magnetic}
    \diffp {b}{t} = \quad 
    r\mathbf{B_0} \cdot\nabla\left(\frac{v}{r}\right). 
\end{equation}

Now let us see what happens to an initially incompressible system with purely torsional perturbations when nonlinear effects are included. We consider a time at which $v_r,b_r,v_z,b_z$ and  $\rho' = 0$ but this time we include all of the nonlinear terms. This gives us,

\begin{equation} 
\begin{split}
\rho_0 \partial_t v_r
  = \quad & - \frac{1}{\mu_0}b_\theta\partial_r b_\theta - \frac{\rho_0 v_\theta^2}{r} - \frac{1}{\mu_0}\frac{b_\theta^2}{r},
\end{split}
\label{eq:vr}
\end{equation}
\begin{equation} 
\begin{split}
\rho_0 \partial_t v_\theta
  = \quad & \frac{1}{\mu_0}\left(\frac{B_{0r}}{r}\partial_r(rb_\theta)+B_{0z}\partial_z b_\theta\right),
\end{split}
\end{equation}
\begin{equation} 
\begin{split}
\rho_0 \partial_t v_z
  = \quad & -\frac{1}{\mu_0} b_\theta\partial_zb_\theta,
\end{split}
\label{eq:vz}
\end{equation}
\
\begin{equation} 
\begin{split}
\partial_t b_r
  = \quad & 0,
\end{split}
\end{equation}
\begin{equation} 
\begin{split}
\partial_t b_\theta 
  = \quad & \partial_z (B_{0z}v_\theta)+\partial_r (B_{0r}v_\theta),
\end{split}
\end{equation}
\begin{equation} 
\begin{split}
\partial_t b_z 
  = \quad & 0,
\end{split}
\end{equation}
\
\begin{equation}
\partial_t \rho' = 0.
\end{equation}

This nonlinear system will quickly evolve so that $v_r$ and $v_z$ are no longer zero which will lead to perturbations in $b_r$ and $b_z$ and cause the system to become compressible. 

Looking at the momentum equations for $v_r$ and $v_z$ \cref{eq:vr,eq:vz} we can see that the nonlinear generation of magnetosonic waves is caused by: the varying magnetic pressure of the Alfv\'en wave, $\nabla(\mathbf{b}^2/2\mu_0)$, also known as the ponderomotive force \citep{Hollweg,Tikhonchuk}; the magnetic tension force $(\mathbf{b}\cdot\nabla)\mathbf{b}/\mu_0$ and centrifugal force, $\rho_0(\mathbf{v}\cdot\nabla)\mathbf{v}$. When we expand these terms we can see how they each appear in the momentum equations \cref{eq:vr,eq:vz},

\begin{equation}
\begin{split}
-\frac{1}{2\mu_0}\nabla \mathbf{b}^2 = -\frac{1}{\mu_0}\mathbf{b}\nabla \mathbf{b}= \mathbf{f_1},
\quad \text{where} \quad \mathbf{f_1} = \begin{pmatrix}-\frac{b_\theta}{\mu_0}
\partial_r b_\theta\\0\\-\frac{b_\theta}{\mu_0}\partial_z b_\theta\end{pmatrix}, 
\end{split}
\end{equation}
\begin{equation} 
\begin{split}
\frac{1}{\mu_0}(\mathbf{b}\cdot\nabla)\mathbf{b} = \mathbf{f_2},
\quad \text{where} \quad \mathbf{f_2} =
\begin{pmatrix}-\frac{1}{\mu_0}\frac{b_\theta^2}
{r}\\0\\0\end{pmatrix},
\end{split}
\end{equation}
\begin{equation} 
\begin{split}
\rho_0(\mathbf{v}\cdot\nabla)\mathbf{v}= \mathbf{f_3},
\quad \text{where} \quad \mathbf{f_3} =\begin{pmatrix}-\rho_0\frac{v_\theta^2}{r}\\0\\0\end{pmatrix}.
\end{split}
\end{equation}

The ponderomotive force is present for the propagation of shear Alfv\'en waves but the magnetic tension and centrifugal forces are unique to torsional Alfv\'en waves. It is these three nonlinear forces that cause the generation of magnetosonic waves which can subsequently interact with the torsional Alfv\'en wave. In \citep{Farahani}, which considers the dynamics of torsional Alfv\'en waves in a uniform magnetic flux tube using the second-order thin flux tube approximation, we see the same nonlinear terms appearing in the derived wave equation. Furthermore \citep{Farahani} also concludes that the nonlinear self-interaction of the Alfv\'en waves is caused by interaction of the waves with nonlinearly induced, compressive perturbations.

In \cref{sec:simulations} we use MHD simulations to explore the effects of these nonlinear forces on the propagation of torsional Alfv\'en waves, the excitation of magnetosonic waves and the generation of shocks.

\section{Simulations}
\label{sec:simulations}

The FORTRAN code \textit{Lare3d} \citep{lare3d} is a Lagrangian remap code that solves the full viscous MHD equations over a 3D staggered Cartesian grid. In Paper I we used our finite difference solver \textit{WiggleWave} to simulate solutions for the linearised governing equations \cref{eq:velocity,eq:magnetic} for the propagation of torsional Alfv\'en waves. By comparing the simulation outputs from \textit{Lare3d} to those from \textit{WiggleWave} we can detect nonlinear effects. The numerical setup that we use for our \textit{Lare3d} and \textit{WiggleWave} simulations is the same setup we used for our \textit{WiggleWave} simulations in Paper I and is briefly described in the following subsection for clarity. 

\subsection{Numerical Setup}
\label{sec:setup}
 
 The equilibrium magnetic field is potential, axisymmetric and has no azimuthal component. It is defined by it's radial and vertical components as,

\begin{align}
B_r = B_0e^{-z/H}J_1(r/H), \qquad 
B_z = B_0e^{-z/H}J_0(r/H),
\end{align}
where $H$ is the magnetic scale height, $J_0$ and $J_1$ are Bessel functions of the first kind and of zero and first order respectively. In \textit{Lare3d} we must define this field using Cartesian magnetic field components which are,

\begin{align}
B_x \quad = \quad & \frac{B_0 x}{r} e^{-z/H}J_1(r/H), \\
B_y \quad = \quad & \frac{B_0 y}{r} e^{-z/H}J_1(r/H), \\
B_z \quad = \quad & B_0e^{-z/H}J_0(r/H),
\end{align}

We also calculate magnetic field coordinates as we need these to define our density structure. The magnetic coordinates $\phi$ and $\psi$ are defined using the radius $r$,

\begin{align}
r \quad = \quad & \sqrt{x^2+y^2}\\
\phi \quad = \quad & -He^{-z/H}J_0(r/H), \\ 
\psi \quad = \quad & re^{-z/H}J_1(r/H).
\label{eq:define_magcoord}
\end{align}

\begin{figure}
 \centering
  \includegraphics[width=0.9\linewidth]{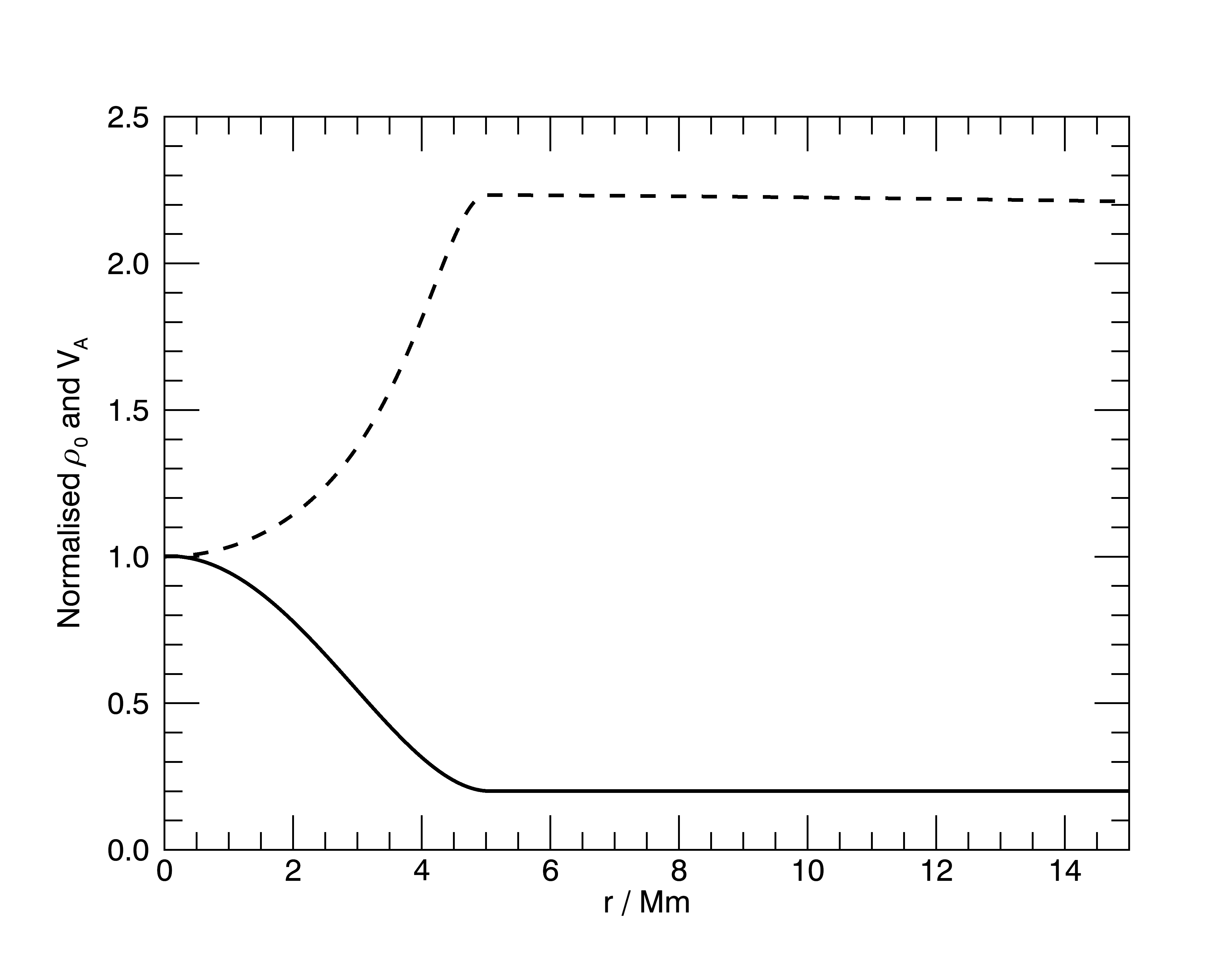}
\caption{A graph showing how the normalised equilibrium density $\rho_0$ (solid line) and normalised Alfv\'en speed $V_A$ (dashed line) change with radius along the lower boundary of the domain i.e. for $z=0$. Note that the density is constant outside the flux tube and is structured within the central flux tube, rather than reaching a plateau.}
\label{fig:rho_va}
\end{figure}

\begin{figure}
\centering
  \includegraphics[width=0.9\linewidth]{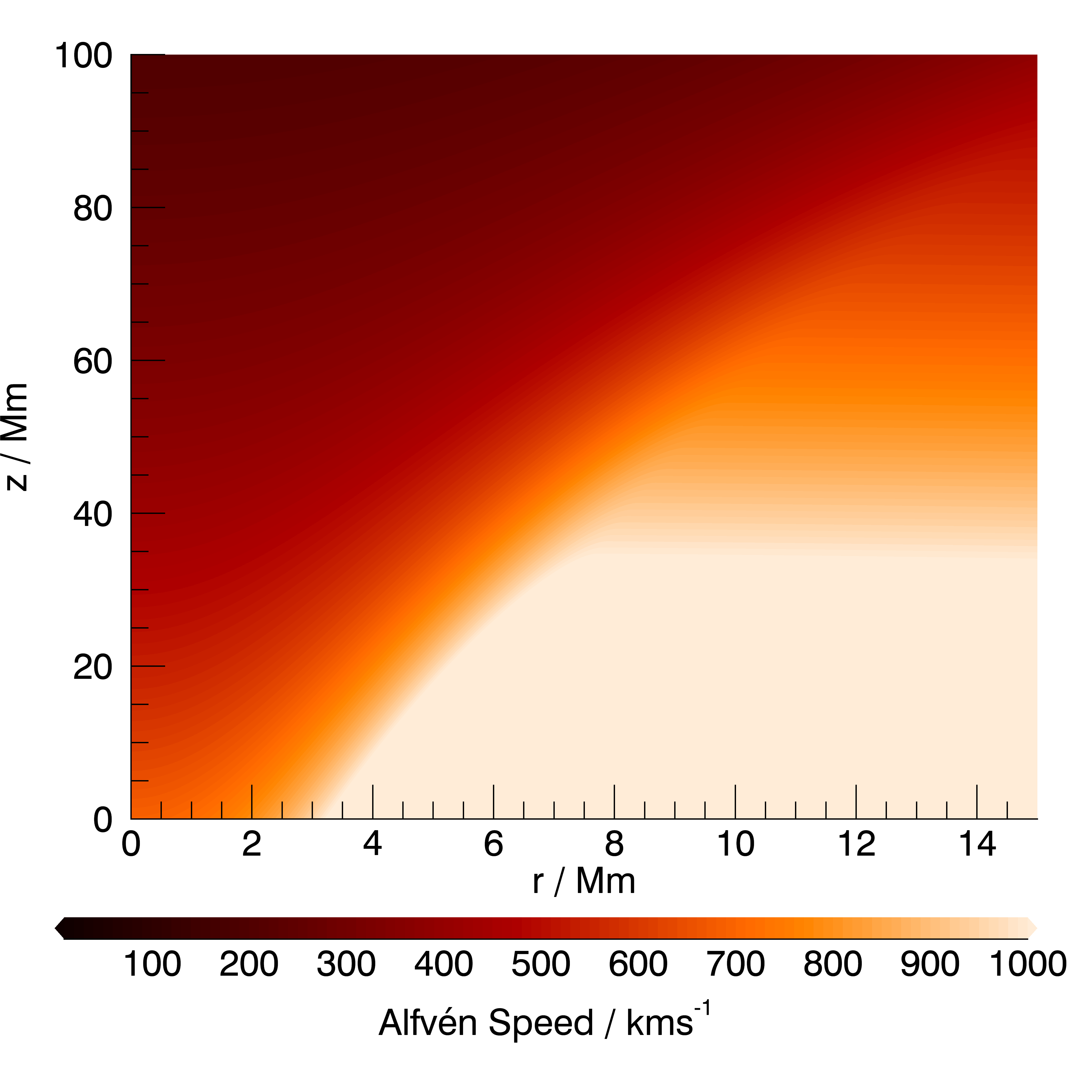}
\caption{A colour contour of the Alfv\'en speed across the computational domain. The azimuthal direction is ignored due to the condition of axisymmetry. The Alfv\'en speed varies both transversely across field lines i.e with $\psi$ and also exponentially with height $z$.}
\label{fig:va}
\end{figure}

We can now define the gravitationally stratified density structure as follows,

\begin{align}
\rho \; = \;  \hat{\rho}(\psi)e^{-z/H_\rho}
 \; = \; \hat{\rho}(\psi)e^{-\alpha z/H},
\end{align}
where the density scale height $H_\rho = k_BT/mg$, $k_B$ is Boltzmann's constant, $T$ is the temperature and is constant, $m \approx 0.6 m_p$, where $m_p$ is the proton mass, $g$ is the solar surface gravity,  $\alpha = H/H_\rho$ and $\hat{\rho}(\psi)$ is an arbitrary function that defines the density variation across magnetic field lines. We use $\hat{\rho}(\psi)$ to define a central higher density tube that is enclosed within the field line defined by $\psi = \psi_b$. The field line $\psi = \psi_b$ intersects the lower boundary at radius $r_0$ which defines the width of the tube. We define $\hat{\rho}(\psi)$ to be,

\begin{align}
\hat{\rho}(\psi) = \frac{\rho_0}{\zeta}
\begin{cases}
    1+(\zeta-1)(1-\psi/\psi_b)^2,& \psi \leq \psi_b\\
    1,              & \psi \geq \psi_b
\end{cases}
\end{align}
where $\zeta$ is the density contrast between the centre and exterior of the magnetic tube $\hat{\rho}(0)/\hat{\rho}(\psi_b)$. The transverse density and Alfv\'en speed profiles along the lower boundary $z=0$ are shown in \cref{fig:rho_va} and the Alfv\'en speed across the whole domain is shown in \cref{fig:va}.


Finally we define our wave driving by prescribing the velocity and magnetic field at the lower boundary, which we approximate as a flat magnetic surface. The wave driving for both $v_\theta$ and $b_\theta$, within the tube boundary $r = r_0$ are given by,

\begin{align}
v \quad = \quad &
    u_0\frac{r}{r_0}\left(1-\left(\frac{r}{r_0}\right)^2\right)
    \sin\left(\omega\left(\frac{z}{V_A}-t\right)\right), \\
b \quad = \quad &
    -u_0\frac{r}{r_0}\left(1-\left(\frac{r}{r_0}\right)^2\right)
    \sin\left(\omega\left(\frac{z}{V_A}-t\right)\right) \sqrt{\mu_0\rho}.   
\end{align}
In \textit{Lare3d} we must define this driving in terms of $v_x,v_y,b_x$ and $b_y$ the equivalent driving is then defined by,

\begin{align}
v_x \quad = \quad &
    -u_0\frac{y}{r_0}\left(1-\left(\frac{r}{r_0}\right)^2\right)
    \sin\left(\omega\left(\frac{z}{V_A}-t\right)\right), \\
v_y \quad = \quad &
        u_0\frac{x}{r_0}\left(1-\left(\frac{r}{r_0}\right)^2\right)
    \sin\left(\omega\left(\frac{z}{V_A}-t\right)\right), \\
b_x \quad = \quad &
    -u_0\frac{y}{r_0}\left(1-\left(\frac{r}{r_0}\right)^2\right)
    \sin\left(\omega\left(\frac{z}{V_A}-t\right)\right), \\
b_y \quad = \quad &
        u_0\frac{x}{r_0}\left(1-\left(\frac{r}{r_0}\right)^2\right)
    \sin\left(\omega\left(\frac{z}{V_A}-t\right)\right).
\end{align}

Considering propagation of Alfv\'en waves in a typical coronal plume or divergent coronal loop structures, we set our characteristic field strength as $B_0 = 0.001$ T, and our characteristic density as $\rho_0 = 1.66 \times 10^{-12}$ kg m\textsuperscript{-3} (which corresponds to a number density of approximately $n_0 = 1\times10^{15}$ m\textsuperscript{-3}). This gives us a characteristic Alfv\'en speed of $V_0 \approx 700$ km s\textsuperscript{-1}. We fix the initial tube radius to be $r_0 = 5$ Mm and the density contrast to be $\zeta = 5$.

For the simulations in this paper we only consider the scenario for which the magnetic scale height is $H = 50$ Mm, the density scale height is $H_\rho = 50$ Mm (corresponding to a coronal temperature of around 1 MK, see \cite{Aschwanden}), the kinematic viscosity is $\nu = 5 \times 10^7$ m\textsuperscript{2}s\textsuperscript{-1} and the period of wave driving is $T = 60$ s. 

For our initial comparison of \textit{Lare3d} and \textit{WiggleWave} simulations we set the amplitude of our Alfv\'en wave driving at the lower boundary using a value of $u_0 = 100$ km s\textsuperscript{-1} this corresponds to a maximum value of about 40 km s\textsuperscript{-1} (approximately 5\% of the local Alfv\'en speed). These values are what we might expect in the corona based on \cite{Banerjee} and \cite{Doyle}.

For our \textit{WiggleWave} simulation the grid resolution used over the domain is set to $500\times2000$, in the $r$ and $z$ directions respectively. \textit{Lare3d} uses a Cartesian grid so we set our resolution to $1000 \times 1000 \times 2000$, in $x$,$y$ and $z$ directions, to allow a direct comparison with the \textit{WiggleWave} simulation. In both \textit{Lare3d} and \textit{WiggleWave} simulations exponential damping regions were included for the upper vertical and outer radial boundaries when necessary, to prevent wave reflection from these boundaries. 

\subsection{Simulation Comparison}
\label{sec:comparison}

The outputs that we will analyse from our \textit{Lare3d} simulations are from a simulation time of only 578 s. The reason for this is that during these simulations the density can dramatically increase, as we will discuss below, this causes the timestep to decrease and slows down the simulation exponentially due to the CFL condition, \citep{kwatra}. 

The outputs from \textit{Lare3d} are the density $\rho$, velocity field components, $v_x,v_y,v_z$, magnetic field components, $b_x,b_y,b_z$ and pressure $P$. When comparing outputs we present only the velocity components of the wave but the magnetic field perturbations have also been checked and display the same nonlinear effects. We can convert the Cartesian velocity components from \textit{Lare3d} into cylindrical equivalents $v_r$ and $v_\theta$ using the formulae,

\begin{alignat}{2}
v_r = \diffp{r}{t} \quad =  \quad & 
\frac{xv_x+yv_y}{\sqrt{x^2+y^2}}
\\
v_\theta = r\diffp{\theta}{t}  \quad =  \quad & 
\frac{xv_y-yv_x}{\sqrt{x^2+y^2}}
\end{alignat}
$v_z$ is unchanged between Cartesian and cylindrical coordinate systems. When comparing outputs we can use the axisymmetry of the outputs to our advantage. By looking at velocities across the plane $y=0$, we can equate $x=r$, $v_x = v_r$ and $v_y = v_\theta$. 

\begin{figure}
 \centering
  \includegraphics[width=0.9\linewidth]{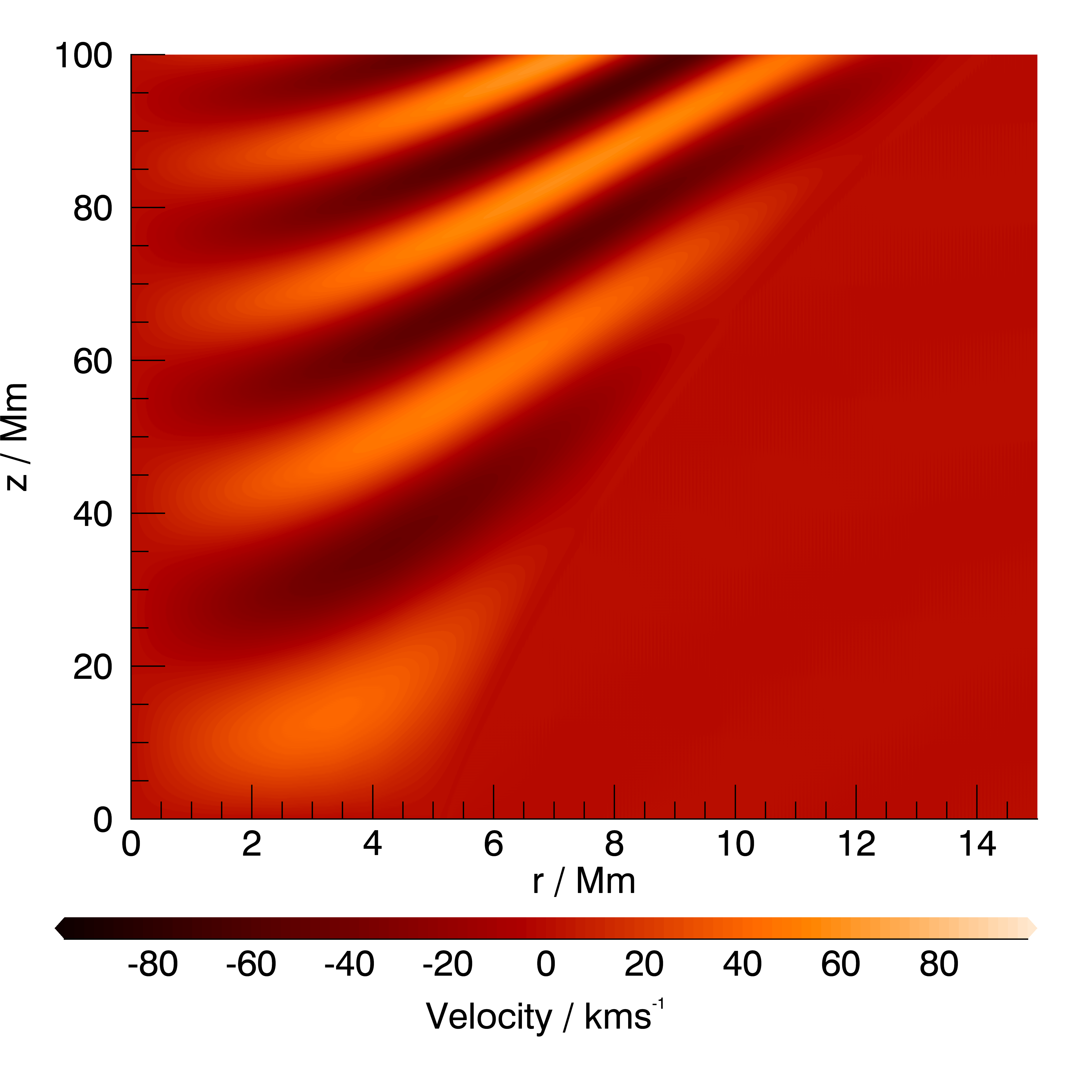}
\caption{A graph of $v_\theta$, over the half plane $y=0$, $x>0$, from the \textit{Wigglewave} simulation of the scenario with $H = 50$ Mm, $H_\rho = 50$ Mm, $\nu = 5 \times 10^7$ m\textsuperscript{2}s\textsuperscript{-1} and $T = 60$ s. We can see the smooth wave fronts of the torsional Alfv\'en wave propagating.}
\label{fig:Nonlin_wiggle}
\end{figure}

\begin{figure}
\centering
  \includegraphics[width=0.9\linewidth]{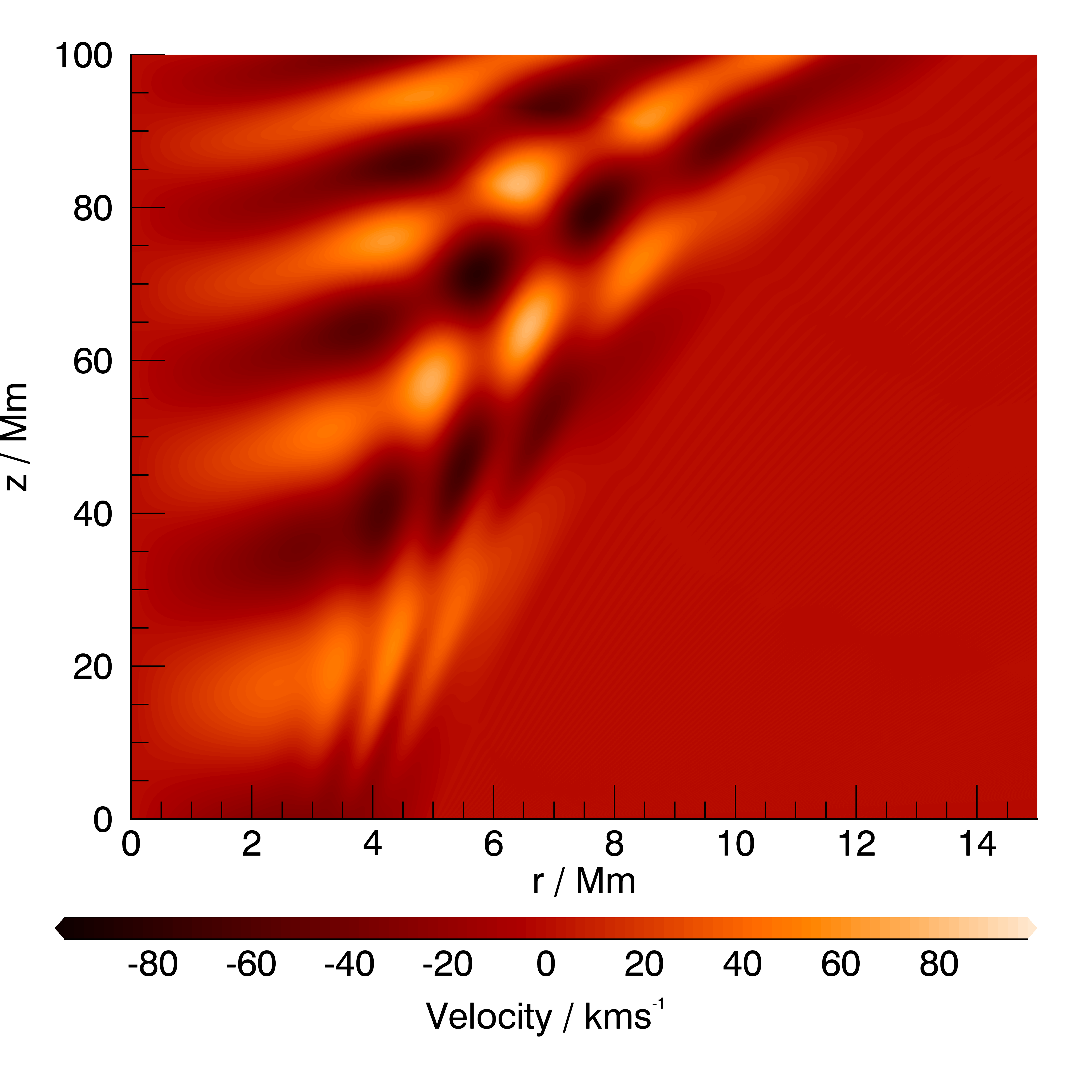}
\caption{A graph of $v_\theta$, over the half plane $y=0$, $x>0$, from the \textit{Lare3d} simulation of the scenario with $H = 50$ Mm, $H_\rho = 50$ Mm, $\nu = 5 \times 10^7$ m\textsuperscript{2}s\textsuperscript{-1} and $T = 60$ s. We can see a fringed pattern superimposed on the wave fronts of the torsional Alfv\'en wave that bends inwards due to refraction that is caused by nonlinear self-interaction.}
\label{fig:Nonlin_lare}
\end{figure}

Contours of $v_\theta$ are shown for the \textit{WiggleWave} and \textit{Lare3d} simulations in \cref{fig:Nonlin_wiggle,fig:Nonlin_lare} respectively. We can see that both simulations show the expected wavefront pattern due to the propagation of the Alfv\'en wave but the \textit{Lare3d} outputs also show a fringed pattern that is superimposed. We will see below that this effect is nonlinear and diminishes at lower amplitudes and that the cause of this effect is nonlinear coupling of the torsional Alfv\'en wave with magnetosonic perturbations.


\begin{figure*}
  \includegraphics[width=\linewidth]{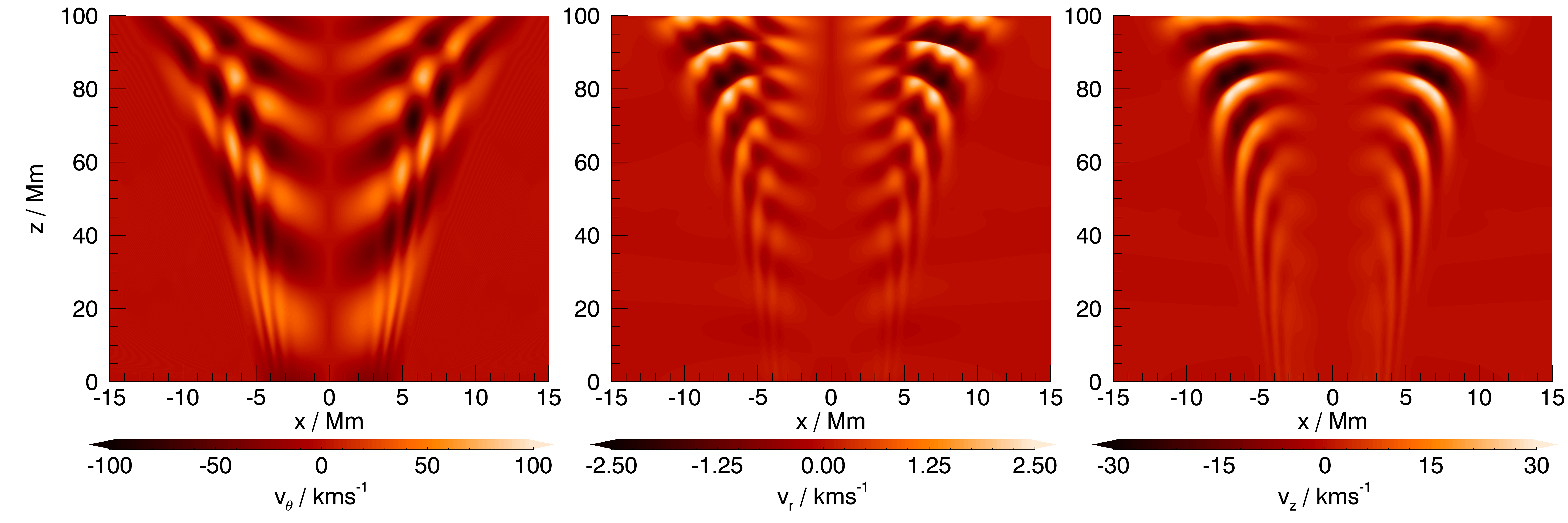}
\caption{Contour plots of the $v_\theta$ (left), $v_r$ (centre) and $v_z$ (right) velocity components, across the plane $y=0$, from the \textit{Lare3d} simulation with an amplitude $u_0$ = 100 km s\textsuperscript{-1}. Note that whilst the contour range for each plot is different the magnitude of $v_z$ perturbations is of a similar order of magnitude to the $v_\theta$ perturbations of the Alfv\'en wave.}
\label{fig:Lare_high}
\end{figure*}

\begin{figure*}
  \includegraphics[width=\linewidth]{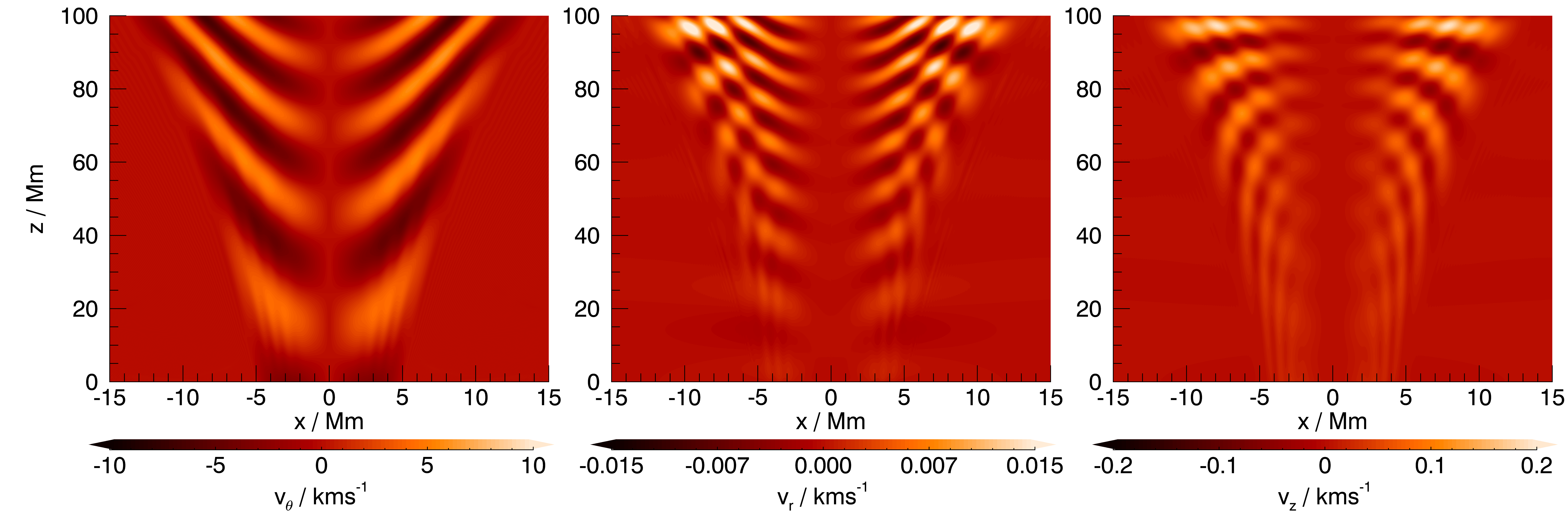}
\caption{Contour plots of the $v_\theta$ (left), $v_r$ (centre) and $v_z$ (right) velocity components, across the plane $y=0$, from the \textit{Lare3d} simulation with an amplitude $u_0$ = 10 km s\textsuperscript{-1}. Note that the contour range for each plot is different and the magnitude of $v_\theta$ Alfv\'en wave perturbations is much greater than $v_r$ and $v_z$ perturbations of the magnetosonic waves.}
\label{fig:Lare_med}
\end{figure*}

These magnetosonic waves are themselves excited nonlinearly by the Alfv\'en waves. The nonlinear excitation of  magnetosonic waves by the phase mixing of Alfv\'en wave is a well known phenomena. It is explained in \cite{Nakariakov1997} that the phase mixing of shear Alfv\'en waves can excite magnetosonic modes. The nonlinear forces responsible for the generation of manetosonic waves are identified in \cref{sec:nonlin}.

The magnetosonic waves can be identified by looking at perturbations to $v_r$ and $v_z$ in the \textit{Lare3d} simulation outputs. Perturbations to $v_r$ and $v_z$ are not calculated in \textit{WiggleWave} as they do not arise in the linear system of equations solved by the code. Perturbations to $v_\theta$, $v_r$ and $v_z$ from our \textit{Lare3d} outputs are shown in \cref{fig:Lare_high}. The presence of perturbations to $v_r$ and $v_z$ is a clear indicator of magnetosonic waves.

Once generated, the magnetosonic waves become trapped inside the central flux tube due to the transverse density structuring.  The enhanced density within the flux tube means that the Alfv\'en velocity is lower inside the flux tube than outside. This causes the waves to refract within the tube which acts a type of waveguide \cite{Roberts,Ofman_hybrid}. This refraction can clearly be seen in \cref{fig:Nonlin_lare,fig:Lare_high}.

To confirm the nonlinear nature of the generated magnetosonic waves we performed two additional \textit{Lare3d} runs with the same parameters but smaller amplitudes. In these two runs we set $u_0 = 10$ km s\textsuperscript{-1} and $u_0 = 1$ km s\textsuperscript{-1}. In \cref{fig:Lare_med} we can see contours of $v_\theta,v_r$ and $v_z$ for the run with $u_0 = 10$ km s\textsuperscript{-1}. Comparing these contours to the ones in \cref{fig:Lare_high} we can see that, besides the amplitudes being smaller, the interference fringes in the $v_\theta$ contour are much less pronounced in the lower amplitude run. Furthermore whilst the amplitude of $v_\theta$ is different by a factor of only 10 between runs, the amplitudes of  $v_r$ and $v_z$ are different by a factor of approximately 100, indicating the nonlinear nature of magnetosonic wave excitement. 

\begin{figure}
  \includegraphics[width=\linewidth]{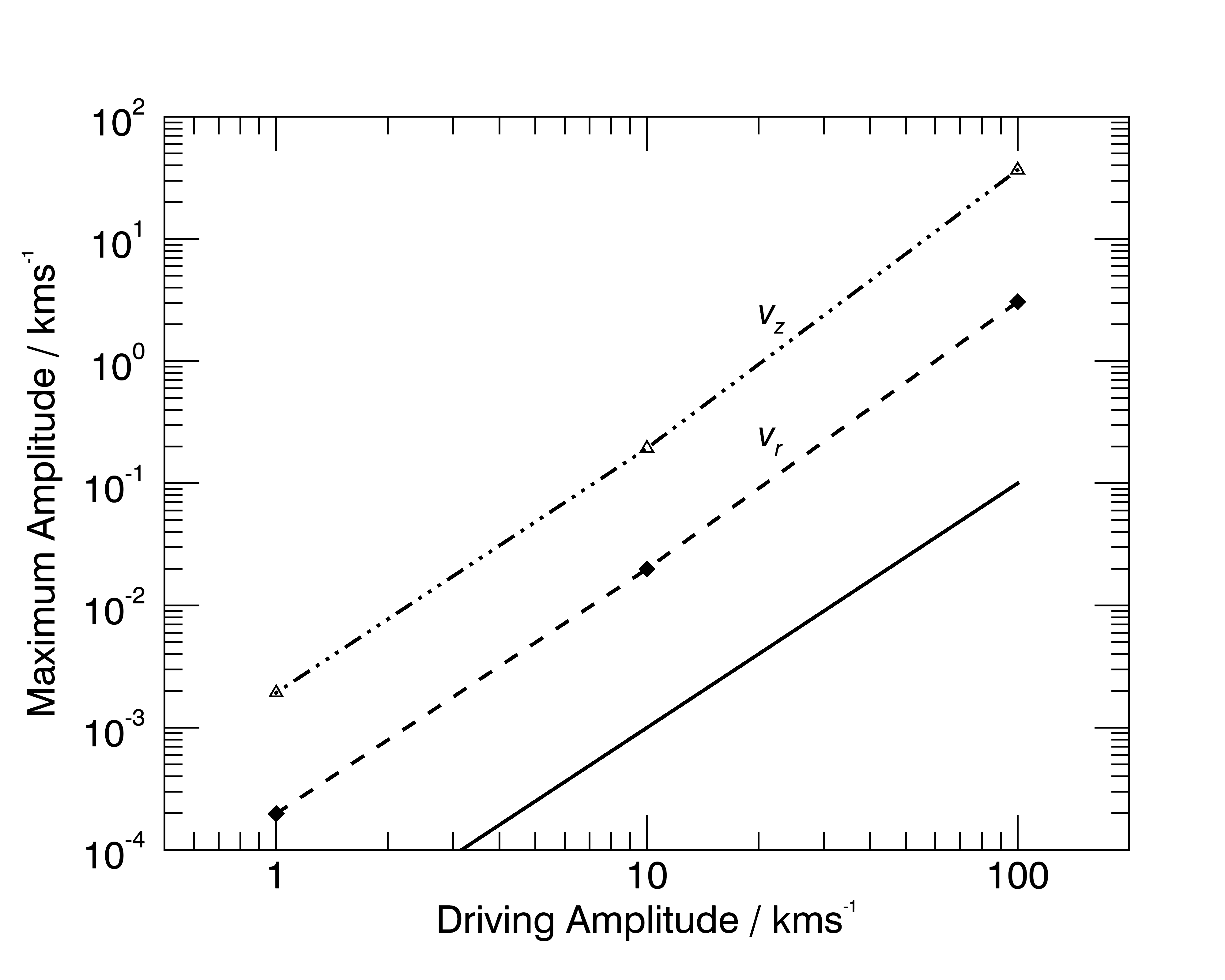}
\caption{A log-log graph showing the maximum amplitudes of $v_r$ (dashed line with diamond data points) and $v_z$ (dot-dashed line with triangular data points), from Lare3d simulation outputs after a simulation time of 578 s. The three simulations used driving amplitudes $u_0 =$ 1, 10 and 100 km s\textsuperscript{-1} as shown on the $x-$ axis. The solid black line is shown for comparison and is an analytic quadratic function of the driving amplitude.}
\label{fig:loglog}
\end{figure}

This point is clarified in \cref{fig:loglog} which shows log-log plots of the maximum values for both $v_r$ and $v_z$ against the driving amplitude $u_0$ across all three simulations. We can see that the slope of these log-log graphs is approximately two, meaning that the amplitude of excited magnetosonic waves scales as roughly the square of the Alfv\'en wave amplitude. Due to the constraints of the CFL condition it is not possible to tell if the magnetosonic waves in our simulations have, or indeed will, reach a point of saturation due to destructive interference as described in \cite{Tsik_strongly,Tsik_weakly}. Nonetheless we might expect this kind of quadratic relationship based on the calculations and simulations for shear Alfv\'en waves made in \cite{Nakariakov1997,Botha} and for torsional Alfv\'en waves in \cite{Shestov}.

The mode conversion to magnetosonic waves indirectly provides another mechanism for the Alfv\'en waves to heat the corona as the generated magnetosonic waves can themselves cause heating through viscous dissipation and shock heating, as discussed in \cite{AntolinShibata}. The presence of the fringed pattern in $v_\theta$ for the \textit{Lare3d} simulations, however, means that we can no longer reliably calculate the wave energy flux from the wave envelope. Despite being unable to measure the damping of the torsional Alfv\'en waves directly from our outputs, we can detect some very relevant physical effects. In addition to detecting the mode conversion and self-interaction of the Alfv\'en wave we can also see shock formation caused by the generated magnetosonic modes as discussed in the following subsection.

\subsection{Shock Formation}
\label{sec:shock}
 
 An important aspect of magnetosonic waves is that they are able to perturb the plasma density. Indeed from \cref{eq:density_pert} in \cref{sec:nonlin} we can see that if $v_r$ and $v_z$ are non-zero then the system is no longer incompressible. It is stated in \citep{Malara_Compressible} that Alfv\'en waves propagating in an compressive nonuniform medium give rise to compressible perturbations that can steepen into shocks that are extremely efficient at dissipating their energy. In \citep{Arber} it is suggested that ponderomotive coupling of Alfv\'en waves to slow modes, which subsequently develop into shocks, is the dominant mechanism that Alfv\'en waves heat the chromosphere. 
 
 Here we will look for evidence of shock formation in our \textit{Lare3d} outputs. Shock formation may provide an alternative and possibly more efficient mechanism for Alfv\'en wave dissipation and one that can potentially be faster than phase mixing \citep{Malara_3d}. To detect compressive perturbations and shocks we consider the density profile from our initial \textit{Lare3d} simulation that uses a wave driving amplitude of $u_0 = 100$ km s\textsuperscript{-1}. Once again we need only consider the density in the plane $y=0$ due to the axisymmetry of our outputs.
 
\begin{figure}
  \includegraphics[width=\linewidth]{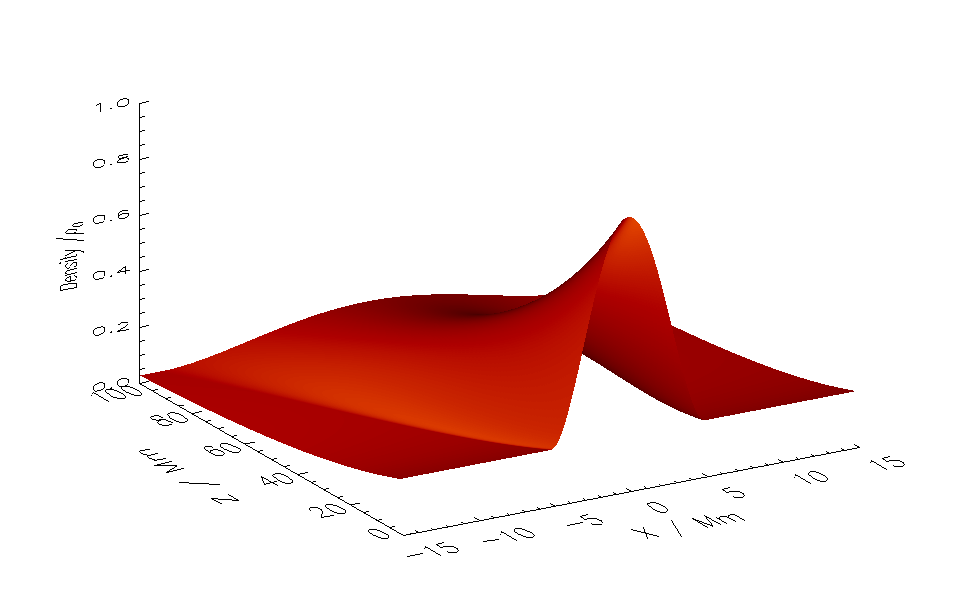}
  \includegraphics[width=\linewidth]{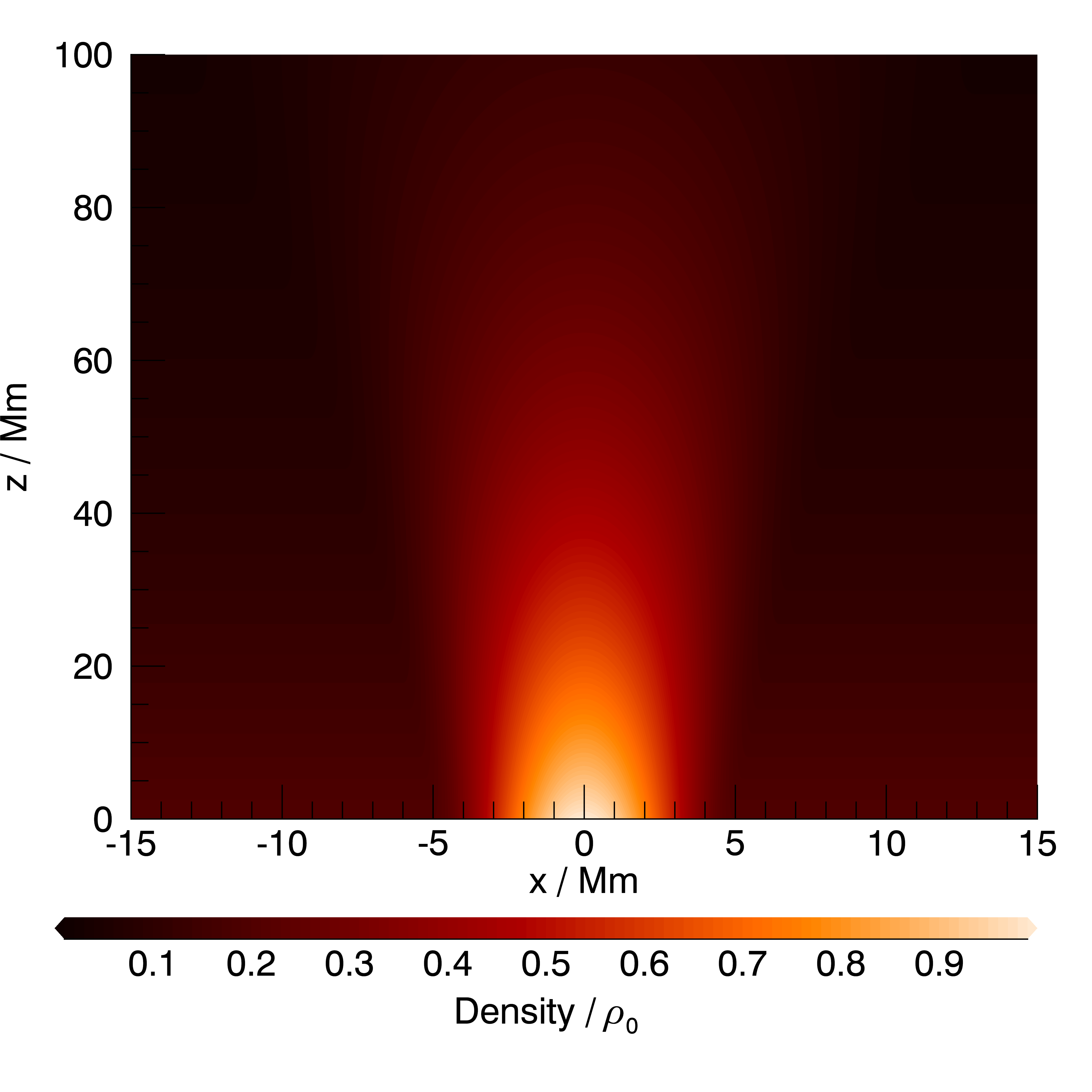}
\caption{A shaded surface (top) and colour contour (bottom) of the initial density across the plane $y=0$ for the \textit{Lare3d} simulation with $u_0$ = 100 km s\textsuperscript{-1}. The densities are shown in terms of the characteristic density $\rho_0 = 1.66 \times 10^{-12}$ kg m\textsuperscript{-3}. The density is higher within the central expanding tube structure and decreases exponentially with height.}
\label{fig:Density_start}
\end{figure}

\begin{figure}
  \includegraphics[width=\linewidth]{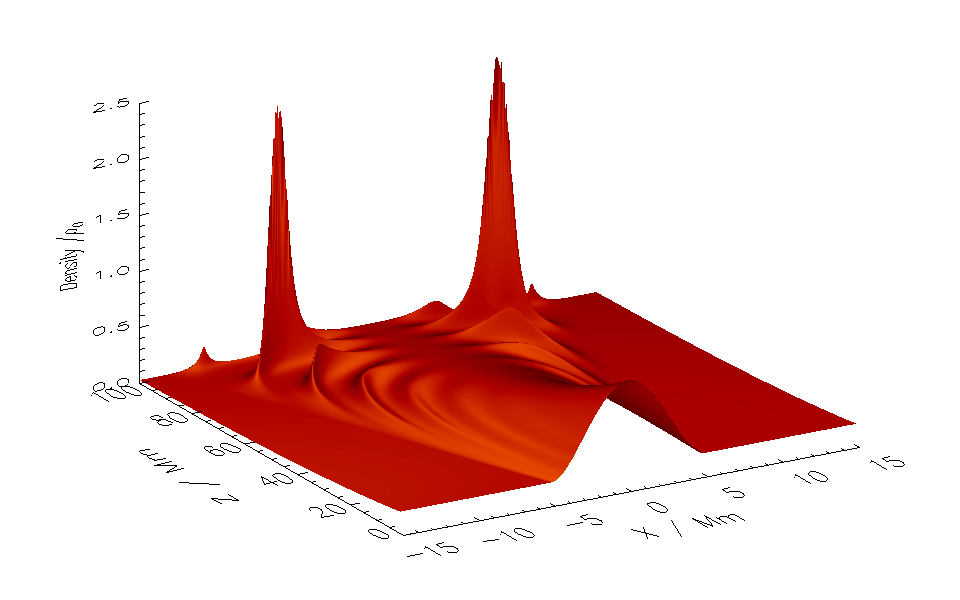}
  \includegraphics[width=\linewidth]{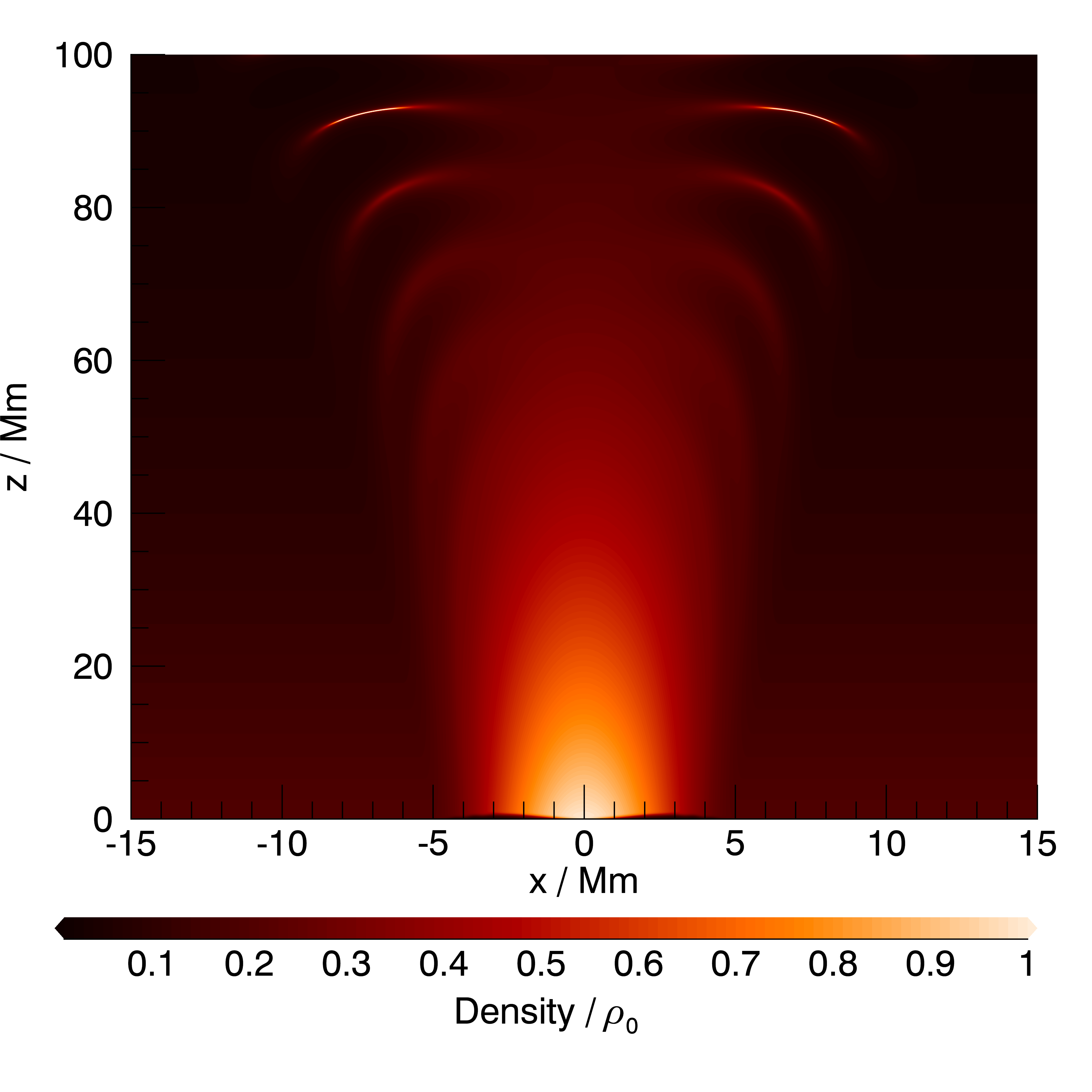}
\caption{A shaded surface (top) and colour contour (bottom) of the density after 578 s across the plane $y=0$ for the \textit{Lare3d} simulation with $u_0$ = 100 km s\textsuperscript{-1}. The densities are shown in terms of the characteristic density $\rho_0 = 1.66 \times 10^{-12}$ kg m\textsuperscript{-3}. There are sharp peaks in the density higher up in the domain reaching densities of almost 2.5 $\rho_0$.}
\label{fig:Density_shock}
\end{figure}

The initial density profile is shown as both as surface plot and a contour in \cref{fig:Density_start} and the density after 578 s of simulation time is shown both as surface plot and a contour in \cref{fig:Density_shock}. We can see from comparing \cref{fig:Density_start,fig:Density_shock} that as the Alfv\'en and magnetosonic waves evolve, compressive waves in the density start to form. These waves are directed along the magnetic field lines rather than across them and can therefore be associated with slow magnetosonic modes. The compressive waves begin to form lower in the corona and increase in amplitude as they propagate higher up. As the waves increase in amplitude they begin to steepen and form shock waves. 

Similar waves are mentioned in \citep{Ofman} that describes quasi-periodic compressional waves that increase in amplitude with height being observed within coronal plumes up to about 140 Mm. The simulations in \citep{Ofman} suggest that these waves are outwardly-propagating slow magnetosonic waves that have become trapped and nonlinearly steepen in the plumes, we may be seeing something similar in our Lare3d simulations.

Furthermore both \citep{Kudoh} and \citep{Cramner} describe the nonlinear coupling of Alfv\'en waves to compressive magnetosonic modes, which then steepen into shocks, as a mechanism for the generation of spicules. This is described as occurring along open magnetic flux tubes such as those found in coronal holes and modelled in our own simulations. Although our simulations take place in the corona rather than the chromosphere it is likely that the same mechanisms are at play.


\subsection{Flow Rates}
\label{sec:flowrate}

One might expect that the formation of shocks from compressive perturbations would result in an outward flow of plasma from the corona. For this reason we decided to calculate the mass flow rate of plasma through each magnetic surface in the direction of the outward magnetic field. We assume that all of the flow takes place within the central flux tube i.e. for $\psi<\psi_b$. We begin by calculating the plasma velocity directed along magnetic field lines as,

\begin{equation}
\frac{\mathbf{v}\cdot\mathbf{B_0}}{B} \quad = \quad (v_rB_r +v_zB_z)/B.
\end{equation}
We then calculate the mass flow rate along magnetic surfaces $\Sigma$ as,

\begin{align}
\label{eq:massflow}
\dot{m}  \; = \; 
\int_{\Sigma_b}{\rho \frac{\mathbf{v}\cdot\mathbf{B_0}}{B}} \; d\Sigma,
\end{align}
where $\Sigma_b$ is the part of the magnetic surface within the central flux tube. Then using the equation for the elementary part of $\Sigma$ which is given in \citep{Ruderman2018} and repeated in Paper I, 

\begin{align}
\label{eq:dsigma}
d\Sigma = \frac{HB_0}{B} d\psi d\theta.
\end{align}
we can simplify our mass flow rate calculation to,

\begin{align}
\label{eq:massflow_calc}
\dot{m}  \; = \; 
2\pi HB_0 \int_{0}^{\psi_b} \rho \frac{\mathbf{v}\cdot\mathbf{B_0}}{B^2} \; d\psi
\end{align}

\begin{figure}
  \includegraphics[width=\linewidth]{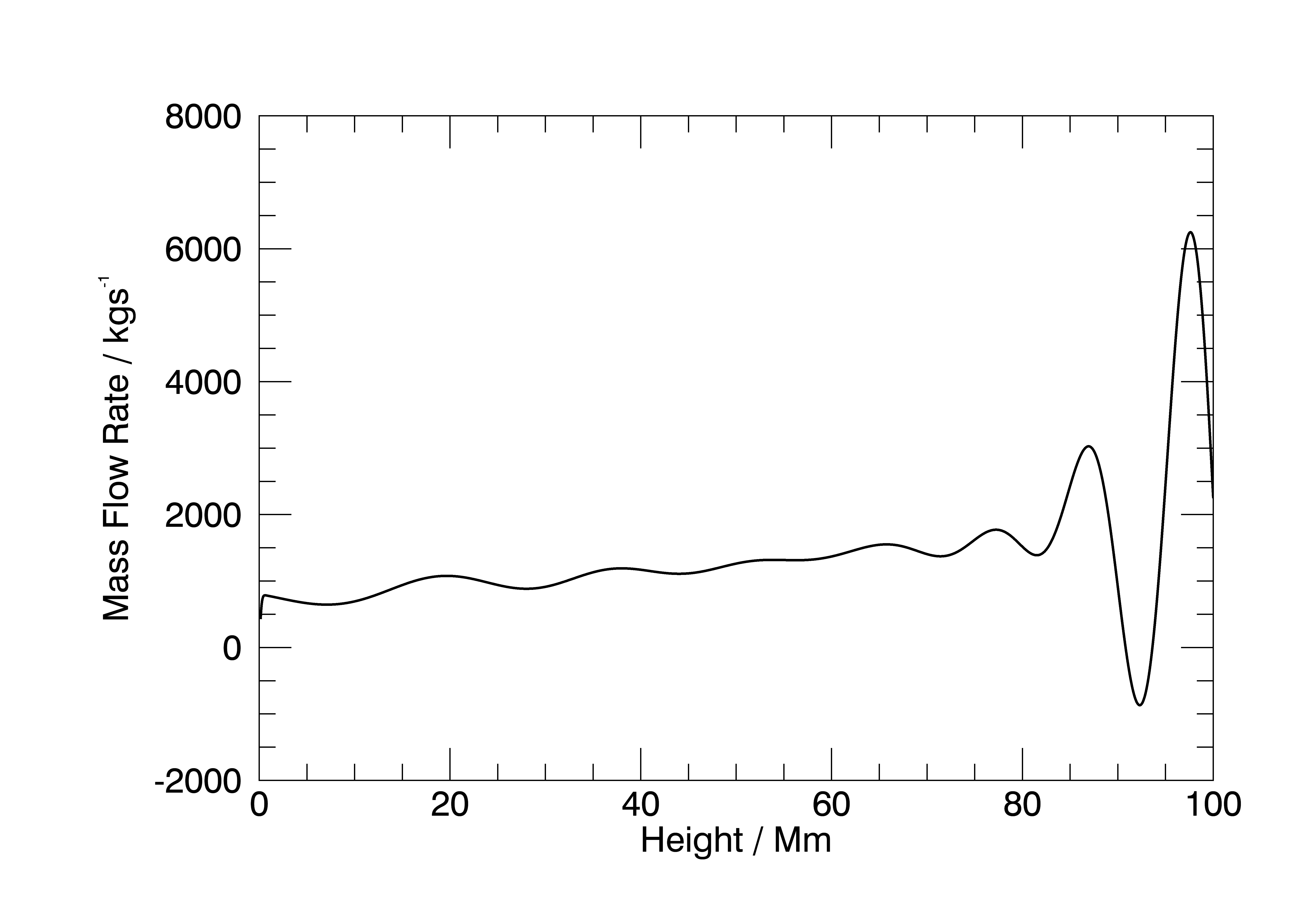}
\caption{Graph of the mass flow rate $\dot{m}$ through magnetic surfaces of increasing height against the height at which each corresponding magnetic surface intersects the $z-$axis for the run with $u_0 = 10$ km s\textsuperscript{-1}.}
\label{fig:peristaltic_med}
\end{figure}

\begin{figure}
  \includegraphics[width=\linewidth]{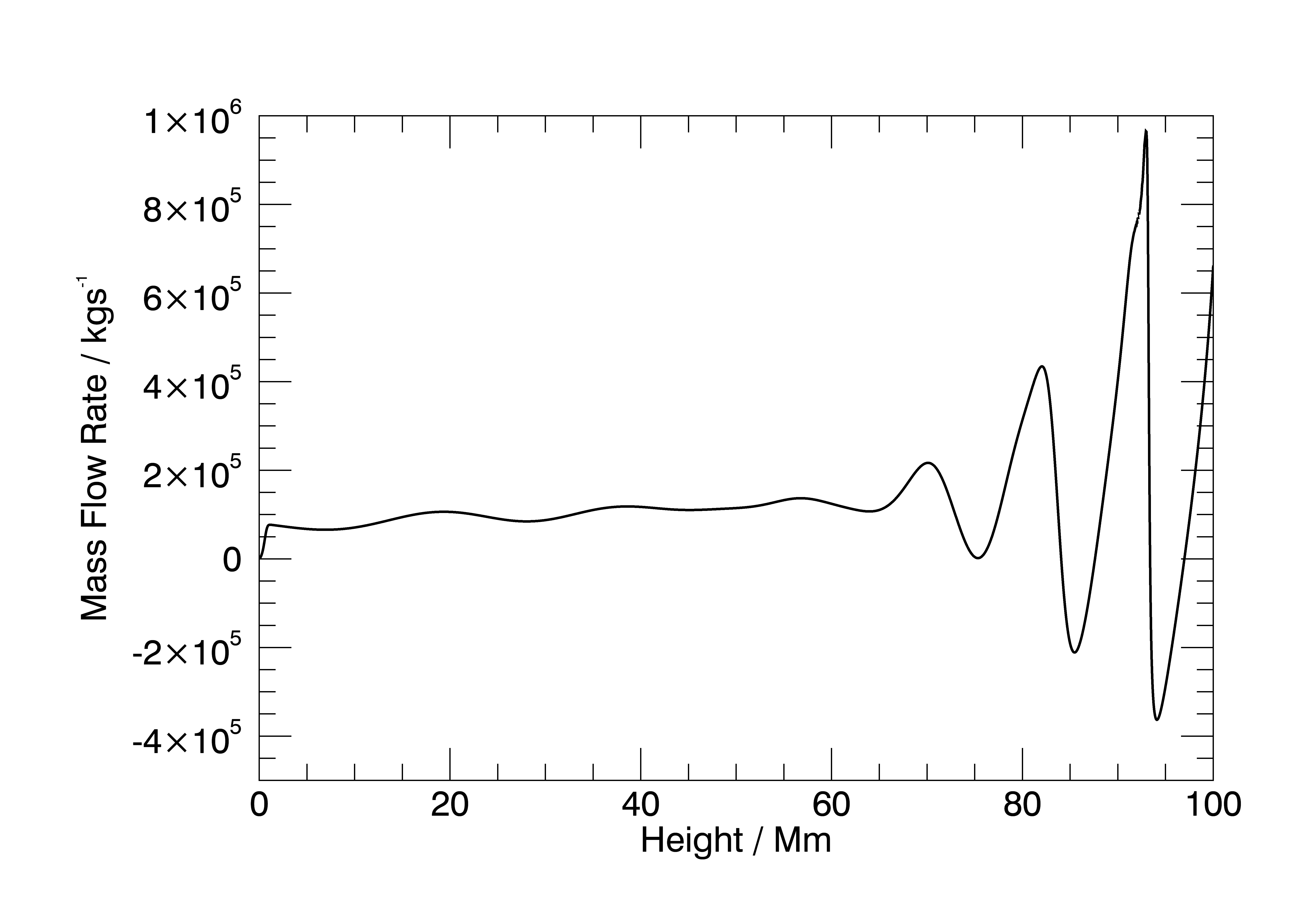}
\caption{Graph of the mass flow rate $\dot{m}$ through magnetic surfaces of increasing height against the height at which each corresponding magnetic surface intersects the $z-$axis for the run with $u_0 = 100$ km s\textsuperscript{-1}.}
\label{fig:peristaltic_high}
\end{figure}

Using this equation we can calculate the mass flow rate through each magnetic surface. In \cref{fig:peristaltic_med,fig:peristaltic_high} we have plotted graphs of the mass flow rate $\dot{m}$ against the height at which each corresponding magnetic surface intersects the $z-$axis. \cref{fig:peristaltic_med} shows the mass flow rate for the run with $u_0 = 10$ km s\textsuperscript{-1} and \cref{fig:peristaltic_high} shows the mass flow rate for the run with $u_0 = 100$ km s\textsuperscript{-1}. 

The mass flow rates shown in \cref{fig:peristaltic_med,fig:peristaltic_high} are for a single point in time and we would expect the flow rates to oscillate in time with the magnetosonic waves. Note that the peak flow rate is generally larger higher up in the domain and so it would not make sense to integrate or average over the height.

We can see that for both simulations the mass flow rate in the direction of the outward magnetic field oscillates but is consistently positive over the lower domain and positive on average over the upper domain. As this nonlinearly induced flow is positive when averaged over an oscillation period it can be considered as an Alfv\'enic wind. This phenomena has also been identified in \citep{Shestov} who perform similar simulations in a straight magnetic flux tube.

The flow in the lower corona is similar to peristaltic flow in compressible viscous fluids \citep{aarts_ooms}, the Alfv\'en wave induces a net flow in the outward direction. In the upper corona the flow shows similarities to the non-Newtonian peristaltic flow, as discussed in \citep{peristaltic}, with the plasma sometimes flowing inwardly, in the opposite direction to the Alfv\'en wave propagation. The effect of the density shocks in the $u_0 = 100$ km s\textsuperscript{-1} simulation can be clearly seen in \cref{fig:peristaltic_high}. The effect is a steepening of the leading edges of the oscillations in the upper corona and an enhancement of the positive peaks in mass flow rate, the combined effect of which is to further increase the net flow rate.


\section{Conclusions}
\label{sec:conclusions}
 
 In this paper we used \textit{Lare3d} \citep{lare3d}, a Lagrangian remap code that solves the full viscous MHD equations over a 3D staggered Cartesian grid, to simulate the propagation of torsional Alfv\'en waves in a divergent and gravitationally stratified solar coronal structure. We compared the simulations outputs to those from corresponding simulations in \textit{WiggleWave} (\url{https://github.com/calboo/Wigglewave}), a finite difference solver we developed to solve the linearised governing equations \cref{eq:velocity,eq:magnetic} directly, in order to detect nonlinear effects.
 
We began in \cref{sec:nonlin} by deriving the full nonlinear equations for the propagation of torsional Alfv\'en waves in a potential axisymmetric magnetic field. We did this to show how the linearised governing equations \cref{eq:velocity,eq:magnetic} are derived and which nonlinear terms are responsible for the generation of non-torsional and compressible perturbations. We showed that the generation of compressible magnetosonic modes by incompressible torsional Alfv\'en is caused by: the varying magnetic pressure of the Alfv\'en wave, $\nabla(\mathbf{b}^2/2\mu_0)$, also known as the ponderomotive force; the magnetic tension force $(\mathbf{b}\cdot\nabla)\mathbf{b}/\mu_0$ and centrifugal force, $\rho_0(\mathbf{v}\cdot\nabla)\mathbf{v}$.

In \cref{sec:simulations} we compared corresponding simulation outputs from \textit{Lare3d} and \textit{WiggleWave}. Comparing the torsional perturbations $v_\theta$ in \cref{fig:Nonlin_wiggle,fig:Nonlin_lare} we saw a fringed pattern superimposed onto the usual wavefront pattern in the \textit{Lare3d} output. We also saw perturbations to $v_r$ and $v_z$ developing in the \textit{Lare3d} outputs that are interpreted as magnetosonic waves excited by coupling of the Alfv\'en wave to magnetosonic modes. The fringed pattern in $v_\theta$ is explained as being caused by the interaction of the induced magnetosonic waves with the Alfv\'en wave.

Following this we ran other simulations in \textit{Lare3d} using the same setup but with different, smaller, amplitudes for wave driving. Comparing the magnitude of magnetosonic waves in these simulations relative to the driving amplitude, as in \cref{fig:loglog}, we showed that the magnetosonic perturbations to $v_r$ and $v_z$ scale as the square of the Alfv\'en wave amplitude, consistent with other research on nonlinear mode coupling \citep{Nakariakov1997,Botha,Shestov}. 

We then turned our attention to the compressive perturbations that are caused by the magnetosonic waves. By inspecting the density outputs from our initial \textit{Lare3d} we can see that as the Alfv\'en and magnetosonic waves evolve compressive waves begin to form lower in the corona and increase in amplitude as they propagate higher up. As the waves increase in amplitude they begin to steepen and form shock waves. 

By considering the mass flow rate through magnetic surfaces we also showed that the net flow is positive in the direction of the magnetic field, showing similarities to peristaltic flow in the lower corona \citep{aarts_ooms} and non-Newtonian peristaltic flow in the upper corona \citep{peristaltic}. Furthermore we showed that the compressive perturbations steepen oscillations in the mass flow rate and enhance the positive peaks.

 To summarise the nonlinear effects identified:
 
 \begin{enumerate}
 
 \item  The nonlinear propagation of a torsional Alfv\'en can excite magnetosonic waves which in turn allow the self-interaction of the Alfv\'en wave, this is shown in \cref{fig:Lare_high}. Mode conversion provides another possible mechanism for the viscous dissipation of Alfv\'en wave energy into the corona. \\
 
  \item  The longitudinal motion of the magnetosonic waves cause compressional waves in the density to form that increases in amplitude with height and steepen into shocks, as seen in \cref{fig:Density_shock}. These shock waves can be extremely efficient at dissipating heat in the corona \citep{Malara_Compressible}. \\
 
 \item  The longitudinal motion of the magnetosonic waves causes a net flow of plasma outwardly along the magnetic field lines as shown in \cref{fig:peristaltic_med,fig:peristaltic_high}. The flow includes oscillations and is not always positive in the higher corona but the positive peaks of the oscillations are enhanced by the presence of the shock waves as seen in \cref{fig:peristaltic_high}. 

 \end{enumerate}

Although we were unable to quantify the rate of torsional Alfv\'en wave damping from our \textit{Lare3d} simulation outputs, due to the interaction of the magnetosonic wave with the Alfv\'en wave, the nonlinear phenomena listed above all provide different mechanisms through which the Alfv\'en wave can dissipate energy. We might therefore expect greater heating rates than predicted from phase mixing alone, due to nonlinear effects in the corona. 

In order to test this theory, however, further simulations must be performed in which viscous coronal heating can be accurately measured. In our current \textit{Lare3d} simulations uniform viscosity is included only as an incompressible term in the momentum equation. If viscosity were included as a term in the energy equation then additional effects may emerge due to increased pressure at heating sites. Furthermore this may provide a method of measuring viscous heating directly.

\section*{Acknowledgements}

C.B. would like to thank UK STFC DISCnet for financial support of his PhD studentship. This research utilized Queen Mary's Apocrita HPC facility, supported by QMUL Research-IT \url{http://doi.org/10.5281/zenodo.438045}.

\section*{Data Availability}

All data used in this study was generated by either our finite difference solver \textit{Wigglewave}, \url{https://github.com/calboo/Wigglewave} or the MHD solver \textit{Lare3d},\citep{lare3d}, \url{https://github.com/Warwick-Plasma/Lare3d}.



\bibliographystyle{mnras}
\bibliography{Enhanced_Phase_Mixing_pt.II} 



\appendix

\section{Nonlinear Cylindrical MHD Equations for Perturbations to an Axisymmetric Magnetostatic Equilibrium}
\label{app1}

The component equations of the cold ($\beta=0$), ideal (zero viscosity, zero resistivity) MHD equations in cylindrical coordinates, ($r,z,\theta$), and with the condition of axisymmetry ($\partial/\partial\theta$) are as follows:

\begin{equation} 
\begin{split}
\rho\left[ \partial_t v_r + v_r \partial_r v_r +v_z\partial_z v_r - \frac{v_\theta^2}{r} \right] = \\
 \quad -\frac{1}{\mu_{0}} \left(\frac{B_\theta}{r}\partial_r(rB_\theta)-B_z\left(\partial_z B_r - \partial_r B_z \right)\right),
  \end{split}
\end{equation}
\begin{equation}
\begin{split}
\rho\left[\partial_t v_\theta + \left(v_r\partial_r v_\theta +v_z\partial_z v_\theta + \frac{v_\theta v_r}{r} \right)\right] = \\
\quad  -\frac{1}{\mu_{0}}\left(-B_r\frac{1}{r}\partial_r(rB_\theta) - B_z\partial_z B_\theta \right),
\end{split}
\end{equation}
\begin{equation}
\begin{split}
\rho\left[ \partial_t v_z + v_r \partial_r v_z +v_z\partial_z v_z \right] = \\
  \quad -\frac{1}{\mu_{0}} \left(B_\theta\partial_z B_\theta-B_r\left( \partial_r B_z - \partial_z B_r \right)\right),
  \end{split}
\end{equation}
\
\begin{equation} 
\partial_t B_r = \quad -\partial_z \left(v_zB_r-v_rB_z\right),
\end{equation}
\begin{equation} 
\partial_t B_\theta
  = \quad \partial_z \left(v_\theta B_z-v_zB_\theta\right) - \partial_r \left(v_rB_\theta-v_\theta B_r\right),
\end{equation}
\begin{equation} 
\partial_t B_z
  = \quad \frac{1}{r}\partial_r \left(r v_zB_r- r v_rB_z\right),
\end{equation}
\
\begin{equation}
\partial_t \rho + \frac{1}{r}\partial_r(r\rho v_r) + \partial_z (\rho v_z)  = 0.
\end{equation}
If we now consider perturbations to an equilibrium state, $\mathbf{v_0} = \mathbf{0}$, $\mathbf{B_0} = (B_{0r},0,B_{0z})$ and $\rho_0 = \rho_0(r,z)$, that take the form, 

\begin{equation}
\begin{split}
\rho \quad = & \quad \rho_0 + \rho', \\
\mathbf{v} \quad = & \quad \mathbf{0} + \mathbf{v},\\
\mathbf{B} \quad = & \quad \mathbf{B_0} + \mathbf{b}, \\
\end{split}
\end{equation}
where our equilibrium magnetic field is a potential field such that,

\begin{equation}
\nabla\times\mathbf{B_0} = \begin{pmatrix}0\\\partial_zB_{0r}-\partial_rB_{0z}\\0\end{pmatrix} =\mathbf{0}.
\end{equation}
then our equations become,

\begin{equation} 
\begin{split}
\rho_0 \partial_t v_r - \frac{B_{0z}}{\mu_0}\left(\partial_zb_r-\partial_rb_z\right)
  = \quad & N_1,
\end{split}
\end{equation}
\begin{equation} 
\begin{split}
\rho_0 \partial_t v_\theta - \frac{1}{\mu_0}\left(\frac{B_{0r}}{r}\partial_r(rb_\theta)+B_{0z}\partial_z b_\theta\right)
  = \quad & N_2,
\end{split}
\end{equation}
\begin{equation} 
\begin{split}
\rho_0 \partial_t v_z - \frac{B_{0r}}{\mu_0}\left(\partial_rb_z-\partial_zb_r\right)
  = \quad & N_3,
\end{split}
\end{equation}
\
\begin{equation} 
\begin{split}
\partial_t b_r + \partial_z (B_{0r}v_z - B_{0z}v_r) 
  = \quad & N_4,
\end{split}
\end{equation}
\begin{equation} 
\begin{split}
\partial_t b_\theta - \left(\partial_z (B_{0z}v_\theta)+\partial_r (B_{0r}v_\theta)\right)
  = \quad & N_5,
\end{split}
\end{equation}
\begin{equation} 
\begin{split}
\partial_t b_z - \frac{1}{r}\partial_r(rB_{0r}v_z-rB_{0z}v_r) 
  = \quad & N_6,
\end{split}
\end{equation}
\
\begin{equation}
\partial_t \rho' + \partial_r\left(r\rho_0 v_r\right) + \partial_z\left(\rho_0 v_z\right) = \quad N_7,
\label{eq:density_pert}
\end{equation}
where all the linear terms are on the LHS and all the nonlinear terms are on the RHS. The nonlinear terms are,

\begin{equation} 
\begin{split}
N_1
  = \quad  -\rho'\partial_t v_r - (\rho_0+\rho')\left[v_r \partial_r v_r +v_z\partial_z v_r + \frac{v_\theta^2}{r} \right] \\
  -\frac{1}{\mu_0} \left(b_\theta\partial_r b_\theta - b_z\left(\partial_zb_r-\partial_rb_z\right)
  + \frac{b_\theta^2}{r}\right),
\end{split}
\end{equation}
\begin{equation} 
\begin{split}
N_2
  = \quad  -\rho'\partial_t v_\theta - (\rho_0+\rho')\left[v_r \partial_r v_\theta +v_z\partial_z v_\theta + \frac{v_\theta v_r}{r} \right] \\
  +\frac{1}{\mu_0} \left(b_r\partial_rb_\theta+b_z\partial_zb_\theta +\frac{b_rb_\theta}{r}\right),
\end{split}
\end{equation}
\begin{equation} 
\begin{split}
N_3
  = \quad  -\rho'\partial_t v_z - (\rho_0+\rho')\left[v_r \partial_r v_z +v_z\partial_z v_z \right] \\
  -\frac{1}{\mu_0} \left(b_\theta\partial_zb_\theta  - b_r\left(\partial_rb_z-\partial_zb_r\right)\right),
\\
\end{split}
\end{equation}
\
\begin{equation} 
\begin{split}
N_4
  = \quad & -\partial_z \left(v_zb_r-v_rb_z\right),
\end{split}
\end{equation}
\begin{equation} 
\begin{split}
N_5
  = \quad & \partial_z \left(v_\theta b_z-v_zb_\theta\right) + \partial_r \left(v_\theta b_r-v_rb_\theta\right),
\end{split}
\end{equation}
\begin{equation} 
\begin{split}
N_6
  = \quad & \frac{1}{r}\partial_r \left(rv_zb_r-rv_rb_z\right),
\end{split}
\end{equation}
\
\begin{equation}
N_7 = \quad - \frac{1}{r}\partial_r\left(r\rho' v_r\right) - \partial_z\left(\rho' v_z\right).
\end{equation}

These are the nonlinear cylindrical equations for perturbations to the axisymmetric magnetostatic equilibrium. 
We can see from these equations that the nonlinear terms, $N_1,N_3,N_4$ and $N_6$ which correspond to the evolution equations for $v_r,v_z,b_r$ and $b_z$, contain terms with either two magnetosonic variables or two Alfv\'en wave variables, indicating that the magnetosonic waves can be excited by Alfv\'en waves and can self-interact. 
In contrast the nonlinear terms, $N_2$ and $N_5$, that correspond to the evolution equations for $v_\theta$ and $b_\theta$, contain only terms with an Alfv\'en wave variable and a magnetosonic variable, indicating that they cannot be excited by magnetosonic waves but can interact with the induced magnetosonic waves.

\section{Linearised equations for torsional Alfv\'en waves in a potential axisymmetric magnetic field}
\label{app2}

Looking at the equations in \cref{app1} we can see that, if the nonlinear terms are ignored, then setting $v_r,b_r,v_z,b_z$ and $\rho' = 0$ results in an incompressible system with only torsional perturbations. The equations become,

\begin{equation} 
\begin{split}
\rho_0 \partial_t v_r
  = \quad & 0,
\end{split}
\end{equation}
\begin{equation} 
\begin{split}
\rho_0 \partial_t v_\theta
  = \quad & \frac{1}{\mu_0}\left(\frac{B_{0r}}{r}\partial_r(rb_\theta)+B_{0z}\partial_z b_\theta\right),
\end{split}
\end{equation}
\begin{equation} 
\begin{split}
\rho_0 \partial_t v_z
  = \quad & 0,
\end{split}
\end{equation}
\
\begin{equation} 
\begin{split}
\partial_t b_r
  = \quad & 0,
\end{split}
\end{equation}
\begin{equation} 
\begin{split}
\partial_t b_\theta 
  = \quad & \partial_z (B_{0z}v_\theta)+\partial_r (B_{0r}v_\theta),
\end{split}
\end{equation}
\begin{equation} 
\begin{split}
\partial_t b_z 
  = \quad & 0,
\end{split}
\end{equation}
\
\begin{equation}
\partial_t \rho' = 0.
\end{equation}
This system remains incompressible and perturbations remain purely torsional, i.e. only in $v_\theta$ and $b_\theta$. The equations for $v_\theta$ and $b_\theta$ can then be simplified as follows,

\begin{equation} 
\begin{split}
\rho_0 \partial_t v_\theta
  = \quad & \frac{1}{\mu_0}\left(\frac{B_{0r}}{r}\partial_r(rb_\theta)+B_{0z}\partial_z b_\theta\right)
\\
  = \quad & \frac{1}{r\mu_0}\left(B_{0r}\partial_r(rb_\theta)+B_{0z}\partial_z (r b_\theta)\right)
\\
  = \quad & \frac{1}{r\mu_0}\left(\mathbf{B_0}\cdot\nabla(rb) \right),
\end{split}
\end{equation}
\
\begin{equation} 
\begin{split}
\partial_t b_\theta 
  = \quad & \partial_z (B_{0z}v_\theta)+\partial_r (B_{0r}v_\theta)
\\
  = \quad & B_{0r}\partial_r v_\theta + B_{0z}\partial_z v_\theta + v_\theta(\partial_r B_{0r} + \partial_z B_{0z})
\\
  = \quad & B_{0r}\partial_r v_\theta + B_{0z}\partial_z v_\theta - \frac{v_\theta B_{0r}}{r}
\\
  = \quad & r\left(\mathbf{B_0}\cdot\nabla\left(\frac{v}{r}\right) \right),
\end{split}
\end{equation}
where we have used Gauss' law for magnetism,

\begin{equation} 
\begin{split}
\nabla\cdot\mathbf{B_0} \quad = \quad & 0, \\ \qquad \Rightarrow \qquad 
\partial_r B_{0r} + \partial_z B_{0z} \quad = \quad & - \frac{B_{0r}}{r},
\end{split}
\end{equation}
 if we reintroduce the viscosity term then we have the same linearised governing equations presented in Paper I,
 
 \begin{equation}
    \rho \diffp {v}{t}  = \quad 
    \frac{1}{r\mu_{0}}\left(\mathbf{B_0}\cdot\nabla(rb)\right)
    + \frac{1}{r}\diffp{}{r}\left(\rho\nu r\diffp{v}{r}\right)
    + \diffp{}{z}\left(\rho\nu\diffp{v}{z}\right),
\end{equation}
\begin{equation}
    \diffp {b}{t} = \quad 
    r\mathbf{B_0} \cdot\nabla\left(\frac{v}{r}\right). 
\end{equation}


\bsp	
\label{lastpage}
\end{document}